\documentclass{elsart}
\usepackage{psfig}
\usepackage{natbib}
\begin{document}
\begin{frontmatter}
\title{Scientific Highlights of the HETE-2 Mission}
\author[uofc]{D. Q. Lamb}, 
\author[mit]{G. R. Ricker}, 
\author[cesr]{J.-L. Atteia}, 
\author[cesr]{C. Barraud}, 
\author[cesr]{M. Boer}, 
\author[inpe]{J. Braga}, 
\author[mit]{N. Butler}, 
\author[goddard]{T. Cline}, 
\author[mit]{G. B. Crew}, 
\author[cesr]{J.-P. Dezalay}, 
\author[uofc]{T. Q. Donaghy}, 
\author[mit]{J. P. Doty}, 
\author[mit]{A. Dullighan}, 
\author[lanl]{E. E. Fenimore}, 
\author[lanl]{M. Galassi}, 
\author[uofc]{C. Graziani}, 
\author[ucb]{K. Hurley}, 
\author[ucb]{J. G. Jernigan}, 
\author[titech]{N. Kawai}, 
\author[mit]{A. Levine}, 
\author[tata]{R. Manchanda}, 
\author[nasda]{M. Matsuoka}, 
\author[mit]{F. Martel}, 
\author[mit]{G. Monnelly}, 
\author[mit]{E. Morgan}, 
\author[cesr]{J.-F. Olive}, 
\author[cndr]{G. Pizzichini}, 
\author[mit]{G. Prigozhin}, 
\author[titech]{T. Sakamoto}, 
\author[naoj]{Y. Shirasaki}, 
\author[titech]{M. Suzuki}, 
\author[miyazaki]{K. Takagishi}, 
\author[riken]{T. Tamagawa}, 
\author[riken]{K. Torii}, 
\author[mit]{R. Vanderspek}, 
\author[cesr]{G. Vedrenne}, 
\author[mit]{J. Villasenor}, 
\author[ucsb]{S. E. Woosley}
\author[miyazaki]{M. Yamauchi} and 
\author[aoyama]{A. Yoshida}

\address[uofc]{Department of Astronomy \& Astrophysics, University of Chicago,
				Chicago, IL 60637, USA}
\address[mit]{MIT Center for Space Research, Cambridge, MA 02139, USA}
\address[cesr]{Centre D'Etude Spatiale des Rayonnements, France}
\address[inpe]{Instituto Nacional de Pesquisas Espaciais, Brazil}
\address[goddard]{NASA Goddard Space Flight Center, Greenbelt, MD 20771, USA}
\address[lanl]{Los Alamos National Laboratory, Los Alamos, NM, USA}
\address[ucb]{UC Berkeley, Space Sciences Laboratory, Berkeley, CA 94720, USA}
\address[titech]{Tokyo Institute of Technology, Tokyo, Japan}
\address[tata]{Department of Astronomy and Astrophysics,
	Tata Institute, Mumbai, 400 005, India}
\address[nasda]{NASDA, Tokyo, Japan}
\address[cndr]{Consiglio Nazionale Delle Ricerche, Italy}
\address[naoj]{National Astronomical Observatory, Tokyo, Japan}
\address[miyazaki]{Faculty of Engineering, Miyazaki
	University, Gakuen Kibanadai Nishi, Miyazaki, 889-2192, Japan}
\address[riken]{Institute of Physical and Chemical Research (RIKEN), Tokyo, Japan}
\address[ucsb]{Department of Astronomy \& Astrophysics, University of California, 
				Santa Cruz, CA 95064, USA}
\address[aoyama]{Aoyama University, Tokyo, Japan}

\vskip 0.5truein

\begin{abstract}
The HETE-2 mission has been highly productive.  It has observed more
than 250 GRBs so far.  It is currently localizing 25 - 30 GRBs per
year, and has localized 43 GRBs to date.  Twenty-one of these
localizations have led to the detection of X-ray, optical, or radio
afterglows, and as of now, 11 of the bursts with afterglows have
redshift determinations.  HETE-2 has also observed more than 45 bursts
from soft gamma-ray repeaters, and more than 700 X-ray bursts.

HETE-2 has confirmed the connection between GRBs and Type Ic
supernovae, a singular achievement and certainly one of the scientific
highlights of the mission so far.  It has provided evidence that the
isotropic-equivalent energies and luminosities of GRBs may be 
correlated with redshift; such a correlation would imply that GRBs and
their progenitors evolve strongly with redshift.  Both of these results
have profound implications for the nature of GRB progenitors and for
the use of GRBs as a probe of cosmology and the early universe.

HETE-2 has placed severe constraints on any X-ray or optical afterglow
of a short GRB.  It has made it possible to explore the previously
unknown behavior optical afterglows at very early times, and has opened
up the era of high-resolution spectroscopy of GRB optical afterglows.
It is also solving the mystery of ``optically dark'' GRBs, and
revealing the nature of X-ray flashes (XRFs).
\end{abstract}
\end{frontmatter}

\section{Introduction}

Gamma-ray bursts (GRBs) are the most brilliant events in the Universe.
They mark the birth of stellar-mass black holes and involve
ultra-relativistic jets traveling at 0.9999 c.  Long regarded as an
exotic enigma, they have taken center stage in high-energy astrophysics
by virtue of the spectacular discoveries of the past six years.  It is
now clear that they also have important applications in many other
areas of astronomy: GRBs mark the moment of ``first light'' in the
universe; they are tracers of the star formation, re-ionization, and
metallicity histories of the universe; and they are laboratories for
studying core-collapse supernovae.

Three major milestones have marked this journey.  In 1992, results from
the Burst and Transient Source Experiment (BATSE) on board the {\it
Compton Gamma-Ray Observatory} ruled out the previous paradigm (in
which GRBs were thought to come from a thick disk of neutron stars in
our own galaxy, the Milky Way), and hinted that the bursts might be
cosmological \citep{meegan1993}.  In 1997, results made possible by
{\it Beppo}SAX \citep{costa1997} decisively determined the distance
scale to long GRBs (showing that they lie at cosmological distances),
and provided circumstantial evidence that long bursts are associated
with the deaths of massive stars [see, e.g., \cite{lamb2000}].  In
2003, results made possible by the High Energy Explorer Satellite 2
(HETE-2) \citep{vanderspek2003a} dramatically confirmed the GRB -- SN
connection and firmly established that long bursts are associated with
Type Ic core collapse supernovae.  Thus we now know that the
progenitors of long GRBs are massive stars.

The HETE-2 mission has been highly productive in addition to achieving
this breakthrough:

\begin{itemize}

\item 
HETE-2 is currently localizing 25 - 30 GRBs per year;

\item 
HETE-2 has accurately and rapidly localized 43 GRBs in 2 1/2 years of
operation (compared to 52 GRBs localized by {\it Beppo}SAX during its
6-year mission); 14 of these have been localized to $< 2$ arcmin
accuracy by the SXC plus WXM.

\item 
21 of these localizations have led to the identification of the X-ray,
optical, or radio afterglow of the burst.

\item  
As of the present time, redshift determinations have been reported for
11 of the  bursts with afterglows (compared to 13 {\it Beppo}SAX bursts
with redshift determinations).

\item 
HETE-2 has detected 16 XRFs so far (compared to 17 by {\it Beppo}SAX).

\item
HETE-2 has observed 25 bursts from the soft gamma-ray repeaters
1806-20 and 1900+14 in the summer of 2001; 2 in the summer of 2002; and
18 so far in 2003.  It has discovered a possible new SGR: 1808-20.

\item
HETE-2 has observed $\sim$ 170 X-ray bursts (XRBs) in the summer of
2001, $>$ 500 in the summer of 2002, and $>$ 150 so far in 2003 from
$\sim$ 20 sources.  (We pointed HETE-2 toward the Galactic plane during
the summer of 2002 and caught a large number of XRBs in order to 
calibrate new SXC flight software.)

\end{itemize}

Fourteen GRBs have been localized by the HETE-2 WXM plus SXC so far. 
Remarkably, all 14 have led to the identification of an X-ray, optical,
infrared, or radio afterglow; and 13 of 14 have led to the
identification of an optical afterglow.  In contrast, only $\approx$
35\% of {Beppo}SAX localizations led to the identification of an
optical afterglow.  

\section{Scientific Highlights of the HETE-2 Mission}

Confirmation of the GRB -- SN connection is a singular achievement and
certainly one of the scientific highlights of the HETE-2 mission. 
Other highlights of the mission include the following:

\begin{itemize}

\item 
HETE-2 made possible rapid follow-up observations of a short GRB,
allowing severe constraints to be placed on the brightness of any
X-ray or optical afterglow.  

\item
The rapid follow-up observations made possible by HETE-2 have opened
the era of high-resolution spectroscopy of optical afterglows (e.g,
GRBs 020813, 021004, and 030329).

\item
Accurate, rapid HETE-2 localizations sent to ground-based robotic
telescopes have made it possible to explore the previously unknown
behavior of optical afterglows in the 3 - 20 hour ``gap'' immediately
following the burst that existed in the {\it Beppo}SAX era.  This has
confirmed the existence of a very bright, distinct phase lasting
$\approx$ 10 minutes.

\item 
HETE-2 is solving the mystery of ``optically dark'' GRBs.  As already
remarked upon, the identification of an optical afterglow for 13 of 14
GRBs localized by the SXC plus WXM instruments on HETE-2 has shown that
very few long GRBs are truly "optically dark."  

\item 
Optical and NIR follow-up observations made possible by HETE-2 have
provided the best case to date of a GRB whose optical afterglow has
been extinguished by dust, and several examples of GRBs with
exceptionally dim optical afterglows.  These GRBs would very likely
have been classified as ``optically dark'' were it not for the
accurate, rapid localizations provided by HETE-2.  

\item 
HETE-2 is revealing the nature of X-ray flashes (XRFs).  Specifically,
HETE-2 has provided strong evidence that the properties of XRFs,
X-ray-rich GRBs, and GRBs form a continuum, and therefore that these
three types of bursts are the same phenomenon.  

\item
HETE-2 results also show that XRFs may provide unique insights into the
nature of GRB jets, the rate of GRBs, and the role of GRBs in Type Ic
supernovae.  In particular, the HETE-2 results provide evidence that
GRB jets are uniform rather than structured.  They also suggest that
the jets are very narrow, and that the rate of GRBs may be much larger
than has been thought.

\end{itemize}

\section{GRB -- SN Connection}

\begin{figure}[t]
\centerline{\psfig{file=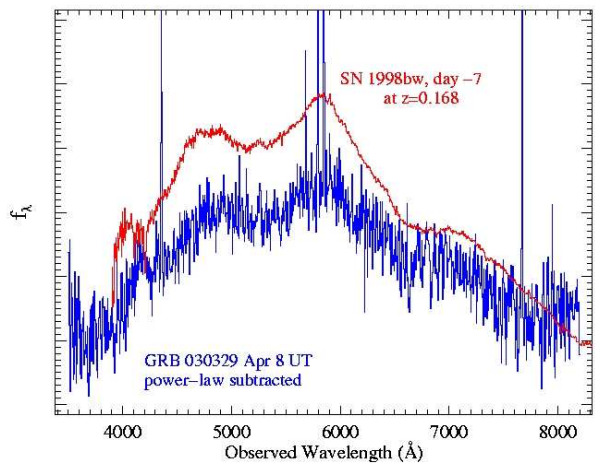,width=0.7 \textwidth}} 
\caption{Comparison of the discovery spectrum of SN 2003dh seen in the 
afterglow of GRB 030329 at 8 days after the burst and the spectrum of
the Type Ic supernova SN 1998bw.  The similarity is striking.  From
\cite{stanek2003}.
\label{fig4}}
\end{figure}

There has been increasing circumstantial and tantalizing direct
evidence in the last few years that GRBs are associated with core
collapse supernovae [see, e.g. \cite{lamb2000}].  The detection and
localization of GRB 030329 by HETE-2 \citep{vanderspek2003a} led to a
dramatic confirmation of the GRB -- SN connection.  GRB 030329 was
among the brightest 1\% of GRBs ever seen.  Its  optical
afterglow was $\sim 12^{\rm th}$ magnitude at 1.5 hours after the burst
\citep{price2003} -- more than 3 magnitudes brighter than the famous
optical afterglow of GRB 990123 at a similar time \citep{akerlof1999}. 
In addition, the burst source and its host galaxy lie very nearby, at a
redshift $z = 0.167$ \citep{greiner2003}.  Given that GRBs typically
occur at $z$ = 1-2, the probability that the source of an observed
burst should be as close as GRB 030329 is one in several thousand.  It
is therefore very unlikely that HETE-2, or even {\it Swift}, will see
another such event.

The fact that GRB 030329 was very nearby made its optical afterglow an
ideal target for attempts to confirm the conjectured association
between GRBs and core collapse SNe.  Astronomers were not disappointed:
about ten days after the burst, the spectral signature of an energetic
Type Ic supernova emerged \citep{stanek2003}.  The supernova has been
designated SN 2003dh.  Figure 1 compares the discovery spectrum of SN
2003dh in the afterglow light curve of GRB 030329 and the spectrum of
the Type Ic supernova SN 1998bw.  The similarity is striking.  The
breadth and the shallowness of the absorption lines in the spectra of
SN 2003dh imply expansion velocities of $\approx$ 36,000 km s$^{-1}$ --
far higher than those seen in typical Type Ic supernovae, and higher
even than those seen in SN 1998bw.  

The clear detection of SN 2003dh in the afterglow of GRB 030329
confirmed decisively the connection between GRBs and core collapse SNe,
and implies that GRBs are a unique laboratory for studying Type Ic core
collapse supernovae.  Confirmation by HETE-2 of the connection between
GRBs and core collapse supernovae has also strengthened the expectation
that GRBs occur out to redshifts $z \sim 20$, and are therefore a
powerful probe of cosmology and the early universe.

\section{Short GRBs}

Assuming that short bursts follow the star-formation rate (as long
bursts are thought to do), \cite{schmidt2001} has shown than the peak
luminosities of short bursts are essentially the same as those of long
bursts.   Otherwise, little is known about the distance scale or the
nature of short GRBs.  {\it Beppo}SAX did not detect any short, hard
GRBs during its 6-year mission, despite extensive efforts.  The rapid
HETE-2 and IPN localizations of GRB 020531 \citep{lamb2003a} made
possible rapid optical ($t$ = 2-3 hours) follow-up observations.  No
optical afterglow was detected
\citep{lamb2002,miceli2002,dullighan2002}.  Chandra follow-up
observations at $t = 5$ days showed that $L_X ({\rm short})/L_x ({\rm
long}) < 0.01 -0.03$ \citep{butler2002}.  These results suggest that
real time or near-real time X-ray follow-up observations of short GRBs
may be vital to unraveling the mystery of short GRBs.

\begin{figure}[t]
\centerline{
\psfig{file=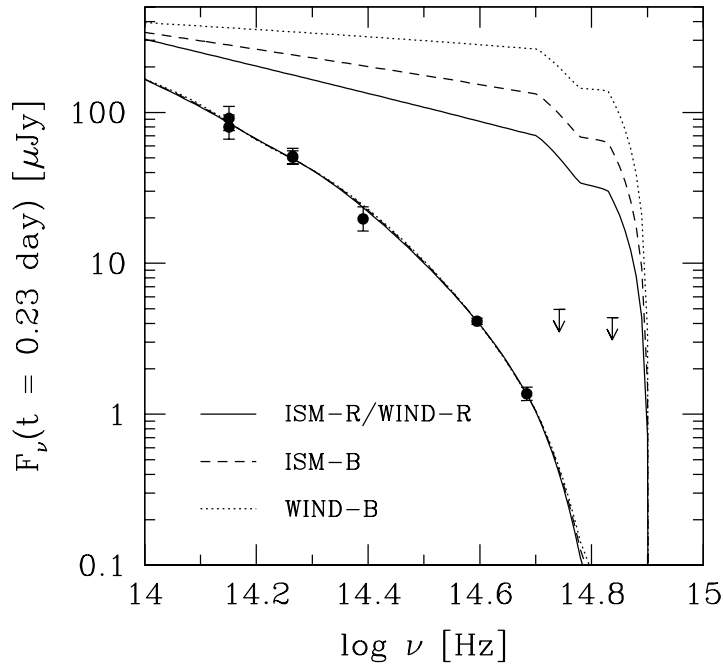,width=0.7 \textwidth}} 
\caption{Left panel: NIR and optical afterglow spectrum of GRB 030115, as
determined from K, H, J, i*, and r* observations.  The curve that goes
through the data points is the best-fit model, assuming extinction by
dust of the four theoretical afterglow spectra labeled "ISM-R, WIND-R,
ISM-B, WIND-B."  The amount of extinction by dust is a sensitive
function of the redshift of the burst.  The redshift of this burst has
not yet been reported; the case shown therefore assumes $z = 3.5$, the
largest redshift allowed by the observations and the one that
attributes the {\it least} amount of extinction by dust.  The amount of
extinction by dust in the optical is still substantial.  From
\cite{lamb2003b}.
\label{fig6}}
\end{figure}

\begin{figure}[t]
\centerline{
\psfig{file=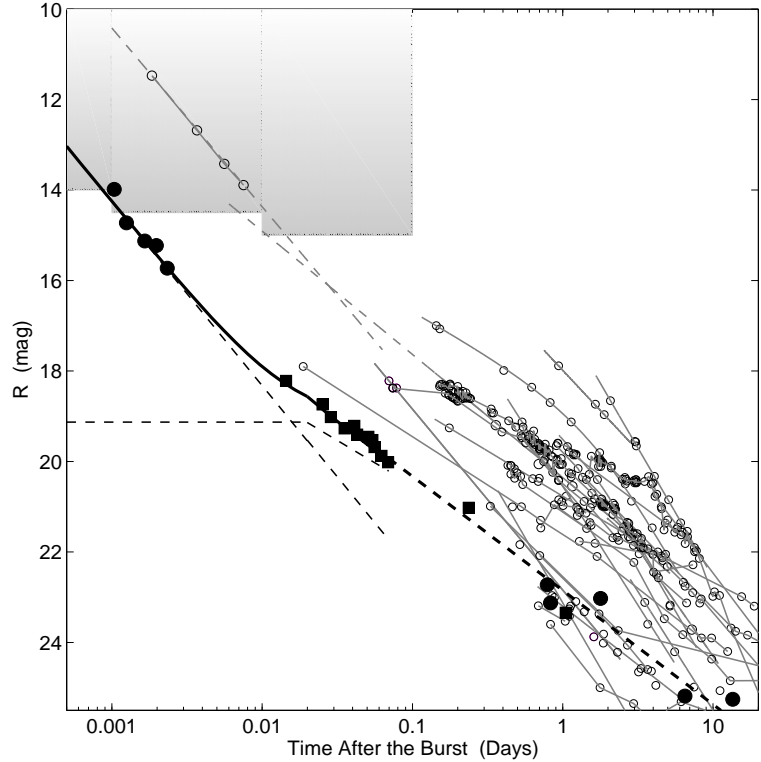,width=0.73 \textwidth}}
\caption{Light curve of the optical afterglow of GRB 021211, compared
to those of other GRBs.  The dashed curve in the upper left-hand corner
of the figure shows the light curve of the optical afterglow of GRB
990123, while the dashed horizontal line in the left-hand middle of the
figure shows the light curve of the optical afterglow of GRB 021004. 
HETE-2 has shown that the optical afterglows of GRBs can exhibit a wide
range of behaviors in the first few hours after the burst.  From
\cite{fox2003b}.
\label{fig7}}
\end{figure}

\section{``Optically Dark'' GRBs}

Only $\approx$ 35\% of {\it Beppo}SAX localizations of GRBs led to the
identification of an optical afterglow.  In contrast, 13 of the 14 GRBs
localized so far by the WXM plus the SXC on HETE-2 have optical
afterglows.  HETE-2 is thus solving the mystery of ``optically dark''
bursts.  

Three explanations of ``optically dark'' GRBs have been widely discussed:

\begin{itemize}

\item 
The optical afterglow is extinguished by dust in the vicinity of the
GRB or in the star-forming region in which the GRB occurs [see, e.g.,
\cite{lamb2000,reichart2002}].

\item
The GRB lies at very high redshift ($z > 5$), and the optical afterglow
is absorbed by neutral hydrogen in the host galaxy and in the
intergalactic medium along the line of sight from the burst to us
\citep{lamb2000b}.

\item
Some GRBs have afterglows that are intrinsically very faint
[see, e.g., \cite{fynbo2001,price2002,lcg2002}].

\end{itemize}

Rapid optical follow-up observations of the HETE-2--localized burst
GRB 030115 \citep{kawai2003} show that the optical afterglow of this
burst is the best case observed to date of a burst whose
optical afterglow is extinguished by dust.  Figure 2 (left panel) shows
the NIR and optical afterglow spectrum of this burst and the best-fit
model, assuming extinction by dust \citep{lamb2003b}.  The amount of
extinction by dust is a sensitive function of the redshift of the
burst.  Since the redshift of this burst has not been reported as yet,
the case shown in Figure 2 (left panel) assumes $z = 3.5$, the largest
redshift allowed by the observations and the one that attributes the
{\it least} amount of extinction by dust.  The amount of extinction by
dust in the optical is still substantial.

Rapid optical follow-up observations
\citep{fox2003b,park2002,li2003,wozniak2002} of the HETE-2--localized
burst GRB 021211 \citep{crew2003} show that the optical afterglow of
this burst is intrinsically much fainter than those observed
previously.  The transition from the reverse shock component
\citep{sari1999} to the forward shock component is clearly visible in
the light curve of the afterglow at about 20 minutes after the burst. 
Figure 2 (right panel) shows the light curve of the afterglow of GRB
021211 compared to those of other GRBs, including a two-component flare
$+$ afterglow fit for the early optical emission from GRB 021211
\citep{fox2003b}.  These observations show that the light curve of the
afterglow of this burst tracks those of GRBs 990123 and 030329, but is
{\it three and six magnitudes fainter} than them, respectively.

This burst would almost certainly have been classified as ``optically
dark'' were it not for its accurate, rapid localization by HETE-2. 
Upper limits or measurements of the optical afterglows of other {\it
Beppo}SAX-- and HETE-2--localized bursts suggest that they too have
afterglows that are very faint [see, e.g., 
\cite{fynbo2001,price2002,lcg2002}].  GRBs with intrinsically faint
afterglows may therefore account for a substantial fraction of bursts
previously classified as ``optically dark.''  

The temporal behavior of the optical afterglow of the HETE-2--localized 
burst GRB 021004 was nearly flat at early times -- a behavior that is
different than any seen previously and that suggests the ``central
engine'' powering the GRB continued to pour out energy long after the
burst itself was over \citep{fox2003a}.  Thus HETE-2 is making it possible
to explore the previously unknown behavior of GRB afterglows  in the
``gap'' in time from the end of the burst to 3 - 20 hours after the
burst that existed in the {\it Beppo}SAX era.

\section{Nature of X-Ray Flashes and X-Ray-Rich GRBs}

As already discussed, the 1, 2, and 6 keV thresholds, and the effective
areas at X-ray energies of the WXM, SXC, and FREGATE instruments make
HETE-2 ideal for detecting and studying XRFs.  Indeed, {\it two-thirds}
of all HETE-2--localized bursts are either ``X-ray-rich'' or XRFs, and
{\it one-third} are XRFs (see Figure 4).\footnote{We define
``X-ray-rich'' GRBs and XRFs as those events for which $\log
[S_X(2-30~{\rm kev})/S_\gamma(30-400~{\rm kev})] > -0.5$ and 0.0,
respectively.}  These events have received increasing attention in the
past several years \citep{heise2000,kippen2002}, but their nature
remains largely unknown.

\begin{figure}[t]
\centerline{
\psfig{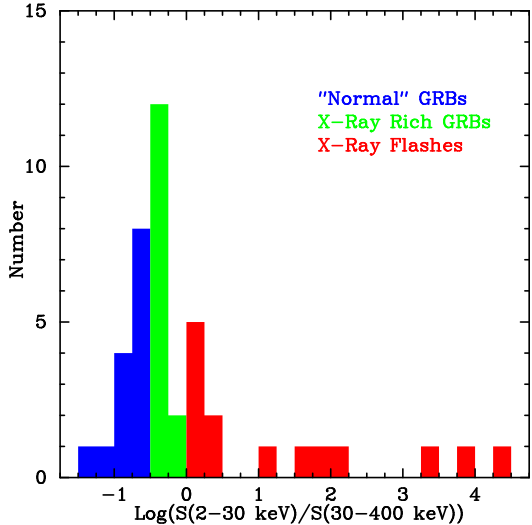}} 
\caption{Hardness histogram for HETE-2 GRBs.  Shown are GRBs
(blue histogram), X-ray-rich GRBs (green histogram), and XRFs (red
histogram).  From \cite{sakamoto2003b}.
\label{fig14}}
\end{figure}

\begin{figure}[t]
\centerline{
\psfig{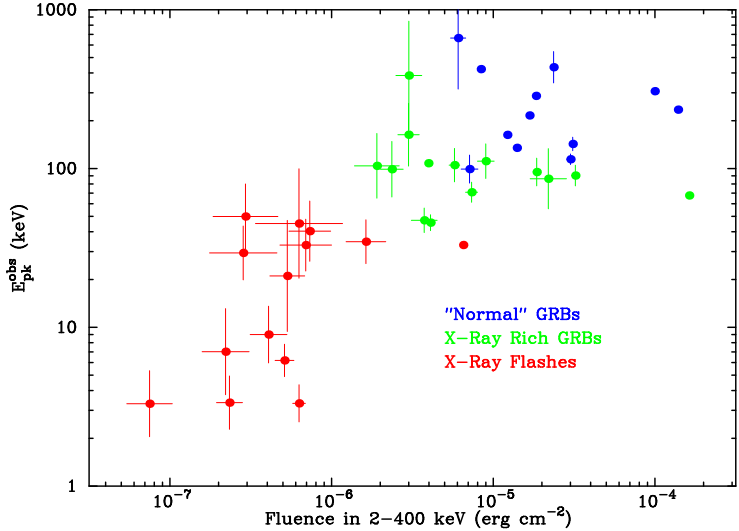}} 
\caption{
Distribution of HETE-2 bursts in the [$S(2-400 {\rm keV}),
E^{\rm obs}_{\rm peak}$]-plane, showing XRFs (red), X-ray-rich GRBs
(green), and GRBs (blue).    From \cite{sakamoto2003b}.
\label{fig14}}
\end{figure}

XRFs have $t_{90}$ durations between 10 and 200 sec and their sky
distribution is consistent with isotropy.  In these respects, XRFs are
similar to ``classical'' GRBs.  A joint analysis of WFC/BATSE spectral
data showed that the low-energy and high-energy photon indices of XRFs
are $-1$ and $\sim -2.5$, respectively, which are similar to those of
GRBs, but that the XRFs had spectral peak energies $E_{\rm peak}^{\rm
obs}$ that were much lower than those of GRBs \citep{kippen2002}. The
only difference between XRFs and GRBs therefore appears to be that XRFs
have lower $E_{\rm peak}^{\rm obs}$ values.  It has therefore been
suggested that XRFs might represent an extension of the GRB population
to bursts with low peak energies.

Clarifying the nature of XRFs and X-ray-rich GRBs, and their connection
to GRBs, could provide a breakthrough in our understanding of the
prompt emission of GRBs.  The spectrum of the HETE-2--localized event
XRF 020903 is exceedingly soft \citep{sakamoto2003a}.  The upper limit
$E^{\rm obs}_{\rm peak} <$ 5 keV (99.7\% confidence level) makes this
event one of the most extreme XRFs seen so far by HETE-2 .  Follow-up
observations made possible by the HETE-2 localization identified the
likely optical afterglow of the XRF \citep{soderberg2002}. Later
observations determined that the optical transient occurred in a
star-forming galaxy at a distance $z = 0.25$
\citep{soderberg2002,chornock2002}; both of these properties are
typical of GRB host galaxies.

Analyzing 42 X-ray-rich GRBs and XRFs seen by FREGATE and/or the WXM
instruments on HETE-2, \cite{sakamoto2003b} find that the XRFs, the
X-ray-rich GRBs, and GRBs form a continuum in the [$S_\gamma(2-400~{\rm
kev}), E^{\rm obs}_{\rm peak}$]-plane (see Figure 5).  This result
strongly suggests that these three kinds of events are the same
phenomenon.

\begin{figure}[t]
\centerline{\psfig{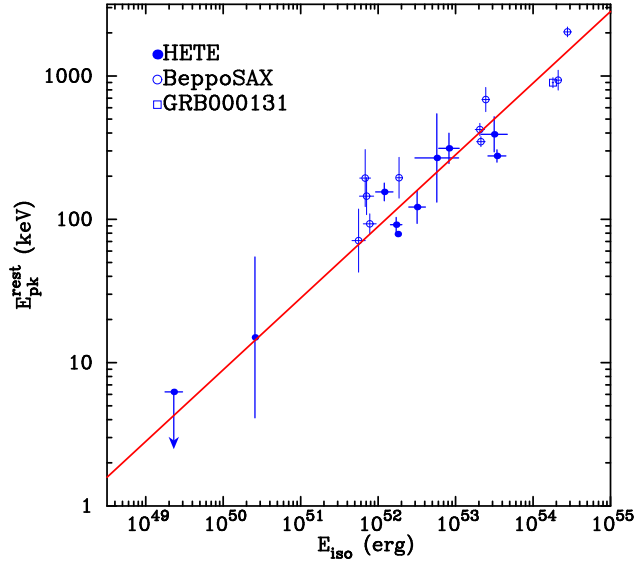}} 
\caption{Distribution of HETE-2 and BeppoSAX bursts in the ($E_{\rm
iso}$,$E_{\rm peak}$)-plane, where $E_{\rm iso}$ and $E_{\rm peak}$ are
the isotropic-equivalent GRB energy and the peak of the GRB spectrum in
the source frame.  The HETE bursts confirm the relation between $E_{\rm
iso}$ and $E_{\rm peak}$ found by Amati et al. (2002), and extend it by
a factor $\sim 300$ in $E_{\rm iso}$.  The bursts with the lowest and
second-lowest values of $E_{\rm iso}$ are XRFs 020903 and 030723. 
From \cite{lamb2003c}.
\label{fig17}}
\end{figure}

Furthermore, \cite{lamb2003c} have placed 9 HETE-2 GRBs with known
redshifts and 2 XRFs with known redshifts or strong redshift constraints
in the ($E_{\rm iso}, E_{\rm peak}$)-plane (see Figure 6).  Here $E_{\rm
iso}$ is the isotropic-equivalent burst energy and $E_{\rm peak}$ is the
energy of the peak of the burst spectrum, measured in the source frame. 
The HETE-2 bursts confirm the relation between $E_{\rm iso}$ and $E_{\rm
peak}$ found by Amati et al. \citep{amati2002} for GRBs and extend it
down in $E_{\rm iso}$ by a factor of 300.  The fact that XRF 020903, one
of the softest events localized by HETE-2 to date, and XRF 030723, the
most recent XRF localized by HETE-2, lie squarely on this relation 
\citep{sakamoto2003a,lamb2003c} provides additional evidence that XRFs
and GRBs are the same phenomenon.  However, additional redshift
determinations are clearly needed for XRFs with 1 keV $< E_{\rm peak} <
30$ keV energy in order to confirm these results.

\section{Conclusions}

The HETE-2 mission has been highly productive.  It has observed more
than 250 GRBs so far.  It is currently localizing 25 - 30 GRBs per
year, and has localized 43 GRBs to date.  Twenty-one of these
localizations have led to the detection of X-ray, optical, or radio
afterglows, and as of now, 11 of the bursts with afterglows have
redshift determinations.  HETE-2 has also observed more than 45 bursts
from soft gamma-ray repeaters, and more than 700 X-ray bursts.

HETE-2 has confirmed the connection between GRBs and Type Ic
supernovae, a singular achievement and certainly one of the scientific
highlights of the mission so far.  It has provided evidence that the
isotropic-equivalent energies and luminosities of GRBs are 
correlated with redshift, implying that GRBs and their progenitors
evolve strongly with redshift.  Both of these results have profound
implications for the nature of GRB progenitors and for the use of GRBs
as a probe of cosmology and the early universe.

HETE-2 has placed severe constraints on any X-ray or optical afterglow
of a short GRB.  It has made it possible to explore the previously
unknown behavior optical afterglows at very early times, and has opened
up the era of high-resolution spectroscopy of GRB optical afterglows.
It is also solving the mystery of ``optically dark'' GRBs, and
revealing the nature of X-ray flashes.

\end{document}